\documentclass[10pt]{article}
\usepackage{fleqn,espcrc1}
\usepackage{graphicx}
\newcommand{\be}{\begin{equation}}
\newcommand{\ee}{\end{equation}}
\newcommand{\bea}{\begin{eqnarray}}
\newcommand{\eea}{\end{eqnarray}}

\def\beq{\begin{equation}}
\def\eeq{\end{equation}}
\def\bea{\begin{eqnarray}}
\def\eea{\end{eqnarray}}
\def\bq{\begin{quote}}
\def\eq{\end{quote}}

\def\ppnp#1#2#3{{\it Prog.~Part.~Nucl.~Phys.~}{\bf #1} (#2) #3}

\def\nnb{\nonumber}
\def\ga{\left(}
\def\dr{\right)}
\def\aga{\left\{}
\def\adr{\right\}}

\def\rar{\rightarrow}
\def\lrar{\Longrightarrow}

\def\nnb{\nonumber}
\def\la{\langle}
\def\ra{\rangle}
\def\nin{\noindent}
\def\ba{\begin{array}}
\def\ea{\end{array}}

\def\als{\alpha_s}

\def\g2{ \la\alpha_s G^2 \ra}
\def\g3{g^3f_{abc}\la G^aG^bG^c \ra}
\def\g4{\la\als^2G^4\ra}

\newcommand{\AmS}{{\protect\the\textfont2
  A\kern-.1667em\lower.5ex\hbox{M}\kern-.125emS}}

\title{\vskip-3cm{\baselineskip14pt
\centerline{\normalsize\hfill PM/00-12}
\centerline{\normalsize\hfill TTP00-04}
\centerline{\normalsize\hfill hep-ph/0003151}
\centerline{\normalsize\hfill March 2000}
}
\vskip.7cm
{Light Hybrid Mesons in QCD}
\vskip.3cm
}

\author{Konstantin Chetyrkin
\footnote{On leave from Institute for Nuclear Research of the Russian
Academy of Sciences, Moscow}
\address{Institut f\"ur theoretische Teilchenphysik, Universit\"at Karlsruhe, 
Kaiserstrasse 12, 76128 Karlsruhe, Germany}
and
Stephan Narison\address{Laboratoire de Physique Math\'ematique et Th\'eorique,
Universit\'e de Montpellier II
Place Eug\`ene Bataillon,
34095 - Montpellier Cedex 05, France and KEK Theory Group, 1-1-Oho, Tsukuba, 
Ibaraki 305,
Japan.
E-mail:
narison@lpm.univ-montp2.fr}
}

\begin{document}

\maketitle

\begin{abstract}
\nin
Including the radiative perturbative corrections and the short distance
tachyonic gluon mass effects
which mimic the ones of UV renormalons, we re-estimate the decay
amplitudes, masses and widths
of light hybrid mesons {}from QCD spectral sum rules. We show that the effects 
are tiny 
and confirm
the previous lowest order results. We discuss the
phenomenological impacts of our results.
\end{abstract}


\section{INTRODUCTION}
\nin
Since the discovery of QCD, it has been emphasized  \cite{GM} that exotic
mesons beyond the
standard octet, exist as a consequence of the non-perturbative aspects of
quantum chromodynamics (QCD). Since the understanding of the nature of the
$\eta'$
\cite{U1}, a large amount of theoretical efforts have been furnished in the
past
and pursued at present for predicting the spectra of the exotics using
different QCD-like models \cite{LIPKIN}
such as the flux tube \cite{REVF} , the bags
\cite{REVB}, the quark \cite{REVP} and constituent gluon 
\cite{REVG} models. In this paper,
we present new developments of the analysis of the hybrid mesons using QCD
spectral sum rules (QSSR)
\cite{SNHYB} \`a la SVZ
\cite{SVZ} (for a review, see e.g.~\cite{SNB}) by including the radiative
perturbative corrections and the short distance tachyonic gluon mass effects
which mimic the ones of UV renormalons
\cite{ZAK,CNZ}. In this sense our results are an update of earlier results. Our
predictions for the masses will be compared with the lattice results
\cite{LATH}
and the recent experimental candidates \cite{CHUNG}.

\section{QCD SPECTRAL SUM RULES (QSSR)}
\subsection*{Description of the method}
\nin
Since its discovery in 1979, QSSR has proved to be a
powerful method for understanding the hadronic properties in terms of the
fundamental QCD parameters such as the QCD coupling $\alpha_s$, the (running)
quark masses and the quark and/or gluon QCD vacuum condensates.
The description of the method has been often discussed in the literature,
where a pedagogical introduction can be, for instance, found in the book
\cite{SNB}. In
practice (like also the lattice), one starts the analysis {}from the
two-point correlator
(standard notations):
\beq
\label{corr}
\Pi^{\mu\nu}_{V/A}(q^2) \equiv i \int d^4x ~e^{iqx} \
\la 0\vert {\cal T}
{\cal O}_{V/A}^\mu (x)
\ga {\cal O}_{V/A}^\nu (0)\dr ^\dagger \vert 0 \ra =-\ga
g^{\mu\nu}q^2-q^\mu q^\nu\dr\Pi^{(1)}_{V/A}(q^2)+q^\mu
q^\nu\Pi^{(0)}_{V/A}(q^2),
\eeq
built {}from the hadronic local currents ${\cal O}^{V/A}_\mu (x)$:
\beq
\label{oper}
{\cal O}_{V}^\mu (x)\equiv :g\bar \psi_i\lambda_a \gamma_\nu \psi_j
G^{\mu\nu}_a: \ ,
\quad {\cal O}_A^\mu (x)\equiv :g\bar \psi_i\lambda_a
\gamma_\nu \gamma_5\psi_j G^{\mu\nu}_a:
\eeq
which select the specific quantum numbers
of the hybrid mesons; A and V refer respectively to the vector and
axial-vector currents.
The invariant $\Pi^{(1)}$ and $\Pi^{(0)}$ refer to the spin one and zero
mesons.
One exploits, in the sum rule approaches, the analyticity property of the
correlator which obeys the well-known K\"allen--Lehmann dispersion relation:
\beq
\Pi_{V/A}^{(1,0)} (q^2) =
\int_{0}^{\infty} \frac{dt}{t-q^2-i\epsilon}
~\frac{1}{\pi}~\mbox{Im} ~ \Pi_{V/A}^{(1,0)}  + ...
\eeq
where ... represent subtraction terms which are
polynomials in the $q^2$-variable. In this way, the $sum~rule$
expresses in a clear way the {\it duality} between the integral involving the
spectral function Im$ \Pi_{V/A}^{(1,0)}(t)$ (which can be measured
experimentally),
and the full correlator $\Pi_{V/A}^{(1,0)}(q^2)$. The latter
can be calculated directly in the
QCD in the Euclidean space-time using perturbation theory (provided  that
$-q^2+m^2$ ($m$ being the quark mass) is much greater than $\Lambda^2$),
and the Wilson
expansion in terms of the increasing dimensions of the quark and/or gluon
condensates which
 simulate the non-perturbative effects of QCD.
\subsection*{Beyond the usual SVZ expansion}
\nin
Using the Operator Product Expansion (OPE) \cite{SVZ}, the two-point
correlator reads for $m=0$:
$$
\Pi^{(1,0)}_{V/A}(q^2)
\simeq \sum_{D=0,2,...}\frac{1}{\ga q^2 \dr^{D/2}}
\sum_{dim O=D} C(q^2,\nu)\la {\cal O}(\nu)\ra~,
$$
where $\nu$ is an arbitrary scale that separates the long- and
short-distance dynamics; $C$ are the Wilson coefficients calculable
in perturbative QCD by means of Feynman diagrams techniques; $\la {\cal
O}(\nu)\ra$
are the quark and/or gluon condensates of dimension $D$.
In the massless quark limit, one may expect
the absence of the terms of dimension 2 due to gauge invariance. However,
it has been
emphasized recently \cite{ZAK} that  the resummation of the large order
terms of the
perturbative series, and the effects of the higher dimension condensates
due e.g. to instantons, can
be mimiced by the effect of a tachyonic gluon mass $\lambda$ which
generates an extra $D=2$ term not present in
the original OPE. Its presence might be understood {}from the analogy with
the short distance linear part
of the QCD potential \footnote{Some evidence of this term is found {}from the
lattice analysis of the static
quark potential \cite{BALI}, though the extraction of the continuum result
needs to be clarified.}. The
strength of this short distance mass has been estimated {}from the
$e^+e^-$ data to be
\cite{CNZ,SNI}:
\beq
\label{lamb}
\frac{\alpha_s}{\pi}\lambda^2\simeq -(0.06\sim 0.07) ~\rm{ GeV}^2,
\eeq
which leads to the value of the square of the (short distance) string tension:
$
\sigma \simeq -\frac{2}{3}{\alpha_s}\lambda^2\simeq [(400\pm 20)~\rm{ MeV}]^2
$
in an (unexpected) good agreement with the lattice result \cite{TEPER} of about
$[(440\pm 38)~\rm{ MeV}]^2$.
In addition to Eq.~(\ref{lamb}), the strengths of the vacuum condensates having
dimensions $D\leq 6$ are also under
good control, namely:
\begin{itemize}
\item $\la\bar ss\ra/\la\bar dd\ra\simeq 0.7\pm 0.2$ {}from the meson
\cite{SNB} and baryon systems
\cite{JAMI2};
\item $\la\alpha_s G^2\ra \simeq (0.07\pm 0.01)~\rm{GeV}^4$ {}from
sum rules of $ e^+e^-\rar I=1~\rm{hadrons}$ \cite{SNI} 
and {heavy quarkonia} \cite{SNH,BB,YNDU},
and {}from the lattice \cite{DIGI2};
\item $g\la\bar{\psi}\lambda_a/2\sigma^{\mu\nu}G^a_{\mu\nu}\psi\ra\simeq
(0.8\pm
0.1)~{\rm GeV}^2\la\bar \psi\psi\ra,$ {}from
the baryons \cite{HEID,JAMI2}, light mesons \cite{OvchPiv1} and
the heavy-light mesons \cite{SNhl};
\item $\alpha_s  \la\bar uu\ra^2\simeq
5.8 \times 10^{-4}~\rm{ GeV}^6$ {}from
$~e^+e^-\rar I=1~ \rm{hadrons}$ \cite{SNI};
\item $g^3\la G^3\ra\approx 1.2~\rm{GeV}^2\la\alpha_s G^2\ra $ 
{}from { dilute gaz instantons}~\cite{NSVZ}.
\end{itemize}
\subsection*{Spectral function}
\nin
In the absence of the complete data,
the spectral function is often parametrized
using the ``na\"{\i}ve" duality ansatz:
\bea
\frac{1}{\pi}~\mbox{Im}  \Pi^{(1,0)}_{V/A}(t)\simeq 2M_H^{4}f_H^2 \delta
(t-M_H^2)+
 \rm{``QCD
~continuum"}
\times \theta(t-t_c)~,
\eea
which has been tested \cite{SNB} using $e^+e^-$ and $\tau$-decay data, to
give a good description of the
spectral integral in the sum rule analysis; $f_H$ 
(an analogue to $f_\pi$) is
the hadron's coupling to the current; $2n$ is the dimension of the
correlator; while $t_c$ is the QCD continuum's threshold.

\subsection*{Form of the sum rules and optimization procedure}
\nin
Among the different sum rules discussed in the literature \cite{SNB}, we
shall be
concerned with the following {Laplace sum rule (LSR)} and its ratio
\cite{SVZ,NR,BB} \footnote{FESR or
$\tau$-like sum rules are complement to the Laplace sum rules and will be
used if necessary, though the final
results are independent on the form of the sum rules used.}:
\beq\label{usr}
{\cal L}^{(1,0)}_n(\tau)
= \int_{0}^{\infty} {dt}~t^n~\mbox{exp}(-t\tau)
~\frac{1}{\pi}~\mbox{Im} \Pi_{V/A}^{(1,0)}(t)~,~~~~~~~~~~{\cal R}_{n}
\equiv -\frac{d}{d\tau} \log {{\cal
L}_n}~,~~~~~~~(n\geq 0) \ .
\eeq
The advantage of the Laplace sum
rules with respect to the previous dispersion relation is the
presence of the exponential weight factor which enhances the
contribution of the lowest resonance and low-energy region
accessible experimentally. For the QCD side, this procedure has
eliminated the ambiguity carried by subtraction constants,
arbitrary polynomial
in $q^2$, and has improved the convergence of
the OPE by the presence of the factorial dumping factor for each
condensates of given dimensions.
The ratio of the sum rules is a useful quantity to work with,
in the determination of the resonance mass, as it is equal to the
meson mass squared, in the usual duality ansatz parametrization.
As one can notice, there are ``a priori" two free external parameters $(\tau,
t_c)$ in the analysis. The optimized result will be (in principle) insensitive
to their variations. In some cases, the $t_c$-stability is not reached due
to the
too na\"{\i}ve parametrization of the spectral function.
In order to restore 
the $t_c$-stability of the
results 
one can either
fix  the
$t_c$-values by the help of FESR (local duality) \cite{Chet,fesr}
or improve the
parametrization of the spectral function by introducing threshold effects
with the help of 
chiral perturbation theory 
The results discussed below satisfy these stability
criteria.

\section{QCD EXPRESSION OF THE TWO-POINT FUNCTION}
\nin
A QCD analysis of the two-point function have been
done in the past by different groups \cite{HYG,HYG1}, where (unfortunately)
the non-trivial QCD expressions were wrong
leading to some controversial predictions \cite{SNB}. The final correct QCD
expression is given in
\cite{HYG2,LNP}. In this paper, we extend the analysis by taking into account
the non-trivial $\alpha_s$ correction and the
effect of the new $1/q^2$ term not taken into account into the SVZ expansion.
The corrected QCD expressions of the correlator are given in
\cite{SNB} to lowest order of perturbative QCD
but including the contributions of the condensates of dimensions lower or
equal than six.
The new terms appearing
in the OPE are presented in the following \footnote{
The results described below  [Eqs. (7) to (9)]
have been obtained with the help of program packages 
GEFICOM (see, e.g. \cite{HarSte98}) 
and MINCER \cite{MINCER} written in FORM \cite{FORM}. 
More details
on the derivation of these results will be published elsewhere.
Note that the results for the NLO radiative corrections  
are derived  in neglecting some possible 
mixings of our operators with those containing more $\gamma$-matrixes
like 
$ 
g\bar \psi_i\lambda_a \gamma_\mu \sigma_{\nu\lambda} \psi_j
G^{\nu\lambda}_a
$
which could in principle mix with ${\cal O}_{V}^\mu$. We expect that effects due 
to
the mixings will be  small.
}:
\begin{itemize}
\item The perturbative QCD expression including the NLO radiative
corrections reads:
\bea
\frac{1}{\pi}\mbox{Im}\Pi_{V/A}^{(1)}(t)_{pert}&=&\frac{\alpha_s}{60\pi^3}t^2
\aga 1+\frac{\alpha_s}{\pi}\Bigg{[}\frac{121}{16}-\frac{257}{360}n_f+\ga
\frac{35}{36}-\frac{n_f}{6}\dr
\log{\frac{\nu^2}{t}}\Bigg{]}\adr\\ \nnb
\frac{1}{\pi}\mbox{Im}\Pi_{V/A}^{(0)}(t)_{pert}&=&\frac{\alpha_s}{120\pi^3}t^2
\aga 1+\frac{\alpha_s}{\pi}\Bigg{[}\frac{1997}{432}-\frac{167}{360}n_f+\ga
\frac{35}{36}-\frac{n_f}{6}\dr
\log{\frac{\nu^2}{t}}\Bigg{]}\adr
\eea
\item The anomalous dimension of the current can be easily deduced to 
be:
\beq
\nu \frac{\mathrm{d}}{\mathrm{d} \nu} {\cal O}_{V}^\mu = 
-\frac{16}{9} \frac{\alpha_s}{\pi} {\cal O}_{V}^\mu
{}.
\label{anom.dim}
\eeq
\item The lowest order correction due to the (short distance)
tachyonic gluon mass reads:
\bea
\frac{1}{\pi}\mbox{Im}\Pi_{V/A}^{(1)}(t)_{\lambda}&=&-\frac{\alpha_s}{60\pi^3}
\frac{35}{4}\lambda^2t\nnb\\
\frac{1}{\pi}\mbox{Im}\Pi_{V/A}^{(0)}(t)_{\lambda}&=&\frac{\alpha_s}{120\pi^3}
\frac{15}{2}\lambda^2t
\eea
\item The (corrected) contributions of the dimension-four and -six terms
reads in the limit $m^2=0$ \cite{SNB}:
\bea
\Pi_{V}^{(1)}(q^2)_{NP}&=&-\frac{1}{9\pi}
\Big{[} \alpha_s\la G^2\ra+8\alpha_s m\la\bar\psi\psi\ra\Big{]}
\log{-\frac{q^2}{\nu^2}}\nnb\\
&&+\frac{1}{q^2}\Big{[}\frac{16\pi}{9}{\alpha_s}
\la\bar \psi\psi\ra^2+\frac{1}{48\pi^2}g^3
\la G^3\ra-\frac{83}{432}\frac{\alpha_s}{\pi}
m g\la\bar\psi G\psi\ra\Big{]}\nnb\\
\Pi_{A}^{(0)}(q^2)_{NP}&=&\Bigg{[}\frac{1}{6\pi}\Big{[} 
\alpha_s\la G^2\ra+8\alpha_s
m\la\bar\psi\psi\ra\Big{]}+\frac{11}{18}
\frac{\alpha_s}{\pi}\frac{1}{q^2}m 
g\la\bar\psi G \psi \ra 
+{\cal O}\ga \frac{1}{q^2}\dr\Bigg{]}\log{-\frac{q^2}{\nu^2}}~,
\eea
where one can notice the miraculous cancellation of the $\log$-coefficient
of the dimension-six condensates for $\Pi_{V}^{(1)}$.
\end{itemize}
\section{PROPERTIES OF LIGHT HYBRIDS}
\subsection*{The $\tilde\rho(1^{-+})$}
\nin
The experimental (resp. theoretical) situation has been
reviewed in \cite{CHUNG} (resp. \cite{SNHYB,LIPKIN}).
The sum rule analysis of the spectrum is based on the
2-point correlator $\Pi(q^2)_{V/A}$ associated to the hybrid currents.
\begin{itemize}
\item One expects, {}from different QCD-like approaches, that the lightest
exotic state is the one with the quantum numbers $1^{-+}$
\footnote{{}From QCD spectral sum rules, we also expect to a good
approximation that the $1^{--}$ is almost
degenerate with the $1^{-+}$.}.
{}From the analysis of the moments ${\cal R}_{0,1}$, we notice that the
effect of the perturbative
corrections (slightly decrease) and of the new dimension-two contribution
(slightly increase) are almost
negligible. This means that perturbation theory expansion in $\als$ 
converges well. 
The main uncertainties come from the value of $t_c$ because the
result does not show $t_c$ stability. The
$\tau$-stability  of ${\cal R}_{0}$ also
disappears if one considers the value of the subtraction constant proposed
in \cite{HYG}:
 \beq q^2\Pi_V^{(1)}|_{q^2=0}\approx \frac{16\pi}{9}\alpha_s
\la \bar\psi\psi\ra^2~,
\eeq
as it would cancel the effect of $\la \bar\psi\psi\ra^2$ appearing in the
OPE. However, we shall check ``a posteriori" that by approximating it with
the sum rule estimated quantity 2$M_{\tilde\rho}^4f_{\tilde\rho}^2$ from
the following Eq. (13), this result is inaccurate and,
therefore, we shall not include this term in our analysis. An independent
measurement of this quantity, e.g. on the lattice is required.
\item Using the different QCD input parameters given previously and the
value $\Lambda=(0.35\pm 0.05)$ GeV,
the positivity ($\equiv t_c\rar
\infty$) of the ${\cal R}_0$ moment leads to the rigorous upper bound:
\beq
M_{\tilde\rho}\leq 1.9~\mbox{GeV}~,
\eeq
which excludes some of range spanned by the quenched lattice estimates 
of $(1.9 \pm 0.2)$ GeV \cite{LATH}.
\item For reasonnable finite values of $t_c\simeq 3.5$ GeV$^{2}$ (beginning of
$\tau$-stability) to 4.5 GeV$^{2}$ as also
fixed by the Finite Energy Sum Rule constraints \cite{LNP,SNB}, we obtain
at the stability point
$\tau\approx (0.5\sim 0.6)$ 
GeV$^{-2}$ of ${\cal R}_0$, the common solution of ${\cal R}_0$ and
${\cal R}_1$:
\beq
M_{\tilde\rho}\approx (1.6\sim 1.7)~\mbox{GeV}~,~~~~~~~~f_{\tilde\rho}\approx
(25\sim 50)~\mbox{MeV}~,~~~~~~~~
(M_{{\tilde\rho}'}\approx\sqrt{t_c})-M_{{\tilde\rho}}\approx 200
~\rm{MeV}~,
\eeq
where the $\tilde\rho'$ is the radial excitation.
One can consider this result as an improvement of the
available sum rule results ranging from 1.4 to 2.1 GeV,
\cite{HYG,HYG1,HYG2,LNP,SNB}. Though, we cannot
absolutely exclude the presence of the $1.4\sim 1.6$  GeV experimental 
candidates \cite{CHUNG}, we expect from
your analysis that this observed state is a hybrid which 
can have a small $\bar qq\bar qq$ component through 
mixing.
\item $\tilde\rho'$-$\tilde\rho$ mass-splitting is much smaller than
$M_{\rho'}-M_\rho\simeq 700$ MeV, and can signal a rich population of
$1^{-+}$ states above 1.6 GeV.
\item The hadronic
widths have been computed in
\cite{LNP,VIRON}. Given our new values of the mass and decay constant, the
updated values are:
\bea
\Gamma(\tilde\rho\rar \rho\pi)&\approx& 274~\mbox{MeV}\ ,
\quad
\Gamma(\tilde\rho\rar \gamma\pi)\approx 3~\mbox{MeV}~,\nnb\\
\Gamma(\tilde\rho\rar \eta'\pi)&\approx& 3~\mbox{MeV}~,~~~~~~~~~~~~~~~~~~~
\Gamma(\tilde\rho\rar \pi\pi,~\bar K K,~\eta_8\eta_8 )
\approx {\cal O}(m^2_q)\ .
\eea
\item One can measure the $SU(3)$ breakings and the mass of the
$\tilde\phi(\bar ss)$
{}from the difference of the ratio of moments, which gives \cite{SNB}:
\beq
M^2_{\tilde\phi}-M^2_{\tilde\rho}\simeq \frac{20}{3}\overline
m^2_s-\frac{160\pi^2}{9}m_s\la
\bar ss\ra \tau \approx 0.3 ~\mbox{GeV}^2~~~~\lrar~~~~ M_{\tilde\phi} 
\approx (1.7\sim 1.8)~\mbox{GeV} \ .
\eeq
The quenched lattice
results are in the range of ($2.0\pm 0.2$) GeV \cite{LATH}, which is slightly 
higher
than our result.
\end{itemize}
\subsection*{The $\tilde\eta(0^{--})$}
\nin
Similar analysis can be done for the pseudoscalar channel. In this case,
the most convenient sum rule to work with is ${\cal R}_1$, which presents both
$\tau$ and $t_c$ stabilities. Using the previous QCD input parameters, 
stabilities are reached for $\tau\simeq 0.12$ GeV$^{-2}$ 
and $t_c\simeq 7.8$ GeV$^2$ showing again that the mass-splitting between the 
radial excitation and the ground state is tiny. At these values 
one obtains \footnote{ In the case $\lambda=0$, i.e. of the ordinary 
SVZ expansion, the stability is obtained
for $\tau\simeq 0.15$ GeV$^{-2}$ at which, one can deduce 
$M_{\tilde \eta}\simeq 3.2$ GeV$\approx\sqrt{t_c}$.}:
\beq
M_{\tilde \eta}\simeq 2.8~{\rm GeV}\approx\sqrt{t_c}~,
\eeq
which we consider as an update of the previous results in \cite{LNP,SNB}.
Again the effects of the correction terms are small.
The relatively higher value of the mass of the $0^{--}$ meson than the one of
the $1^{-+}$, is mainly due to the relative strength of the perturbative and
non-perturbative terms.

\section{CONCLUSIONS}
\nin
There are some progresses in the long run study and experimental search for
the exotics. Before some
definite conclusions, one still needs improvments of the present data, and
some improved lattice unquenched
estimates which should complement the QCD spectral sum rule
(QSSR) results. In this paper we have updated
previous sum rule analysis
of the light hybrids \cite{HYG,HYG1,HYG2,LNP,SNB} by including the perturbative
radiative corrections and the new effect
due to the (short distance) tachyonic gluon mass not included in the original 
SVZ 
expansion.
However, these effects are negligible
which are reassuring for the validity of the approximation used. Our
result which is $M_{\tilde\rho}\approx (1.6\sim 1.7)~\mbox{GeV}$ can be 
renconciled with 
the
existence of the
$1^{-+}$ states at $(1.4\sim 1.6)$ GeV seen recently in hadronic machines (BNL 
and
Crystal Barrel) \cite{CHUNG}, but in the same
time predicts the existence of a
$1^{--}$ hybrid almost degenerate with the $1^{-+}$, and which could
manifest in $e^+e^-\rar$ hadrons by mixing
with the radial excitations of the $\rho$ and $\omega$ mesons. In addition, 
the relatively low value of the continuum threshold indicates that we expect 
a rich population of (axial-) vector hybrids in the region above 1.8 GeV.
\\
In our analysis, the
$0^{-}$ mass is about 2.8 GeV, which is in the range of the different
charmonium states, such that it could mix with these charmonium states 
as well. Moreover, the small splitting between the continuum threshold 
and the lowest ground state indicates that rich population of pseudoscalar
hybrids is expected in the 3 GeV mass range.\\
Light hybrid mesons ($1^{--}$ and $0^{-+}$) might be (partly) responsible of the 
anomalous behaviour of the $e^+e^-\rar$ hadrons cross section observed in the 
region 
below 4 GeV.
\section*{Acknowledgements}
\nin
We are  grateful to Alexei~Pivovarov for
his collaboration on early stage of the present work as well as for 
numerous discussions and a good advice.
K. Ch.  would like to thank Matthias~Steinhauser for his
help with prgramming Feynman rules  for hybrid currents;
his work  was supported by DFG under Contract Ku 502/8-1. 
S.N. thanks the KEK Theory Group and K. Hagiwara for their hospitality.
\vfill\eject

\end{document}